# Photonic Crystal Defect Nanocavities Based on Monocrystalline Yttrium Iron Garnet

Kota Taniguchi*, Siyuan Gao, Tatsuya Kitai, Takeru Yambe, Daisuke Sato, Hironobu Yoshimi, Satoshi Iwamoto, Yasutomo Ota*

K. Taniguchi, S. Gao, T. Kitai, T. Yambe, D. Sato, H. Yoshimi
Department of Applied Physics and Physico-Informatics, Keio University, Yokohama, Kanagawa 223-8522, Japan
E-mail: kota-t10@keio.jp

S. Iwamoto
Institute of Industrial Science, the University of Tokyo, Meguro, Tokyo 153-8505, Japan
Research Center for Advanced Science and Technology, the University of Tokyo, Meguro, Tokyo 153-8904, Japan

Y. Ota
Department of Applied Physics and Physico-Informatics, Keio University, Yokohama, Kanagawa 223-8522, Japan
E-mail: ota@appi.keio.ac.jp

Funding: JST FOREST (JPMJFR213F), JST CREST (JPMJCR19T1) and KAKENHI (25K01697, 24K17582), Iketani Foundation, Nippon Sheet Glass Foundation, JST SPRING Program (JPMJSP2123).

**Abstract text.**

Monocrystalline yttrium iron garnet (YIG) is a key material for magneto optics and quantum magnonics owing to its high optical transparency, large magneto-optical (MO) effects at room temperature, and exceptionally-long spin coherence. While rich MO phenomena have been demonstrated in the microwave regime, extending these concepts to technologically important telecommunication wavelengths remains challenging due to the difficulty of fabricating high-quality YIG nanostructures. Here, we demonstrate photonic crystal (PhC) defect nanocavities based on monocrystalline Bi:YIG by developing a YIG-on-insulator platform and high-precision YIG nanopatterning. The fabricated nanocavities exhibit cavity resonances around λ = 1500 nm with $Q$ factors up to ~1,800 and a mode volume of $V \sim 1.1(\lambda/n)^3$, corresponding to $Q/V$ reaching $\sim 10^3$. The measured $Q$ factor is primarily governed by intentionally introduced lattice modulations for out-of-plane light coupling, suggesting that further optimization of the cavity geometry and measurement configuration could yield an order-of-magnitude improvement of the experimental $Q$ factor. YIG-based PhC nanocavities provide a platform for strongly confined light–magnetism interactions in the optical regime, opening pathways toward downsized nonreciprocal photonic devices and enhanced photon–magnon coupling.

## 1. Introduction

Monocrystalline yttrium iron garnet (YIG) is a unique material that exhibits high optical transparency, strong magneto-optical (MO) effects at room temperature,[1] and exceptionally-long spin coherence.[2] These remarkable properties have enabled a wide range of applications, including nonreciprocal photonics[3] and microwave[4] devices, and quantum magnonics.[5] In the microwave regime, MO interactions in YIG are rather strong, leading to a variety of intriguing demonstrations, ranging from nonreciprocal beam steering,[6] topologically protected unidirectional light transport,[7] to strong photon–magnon coupling.[8] However, realizing these concepts at the technologically-important telecommunication wavelengths has so far been severely limited due to inherently weak MO interactions in the optical regime.

Optical confinement by photonic micro/nanostructures has long been a primary approach to enhance MO interactions in YIG in the optical domain. Early attempts employed one-dimensional YIG-based photonic crystals (PhCs) formed by material deposition,[9] which, however, sacrificed the material's crystallinity. Plasmonic structures have also been examined.[10–12] However, ohmic loss in metals limits the achievable $Q$ factor of optical resonance. More recently, all-dielectric metasurfaces have emerged as low-loss optical confinement structures for YIG. However, experimentally reported $Q$ factors of the all-dielectric YIG metasurfaces have been limited to only around 100, largely hampered by the difficulty in developing YIG photonic nanostructures.[13–15] Micro-sized optical spheres and waveguide-based Fabry-Perot resonators based on YIG have also been explored as high $Q$ resonators.[16–19] While these structures support high $Q$ factors exceeding $10^5$, their confinement volumes are large, thus offering only moderate enhancement of light-matter interactions. Lowering mode volumes for both photons and magnons while keeping low damping rates is rather important for optomagnonics applications,[20] such as magnon-based quantum transducers,[21] magnon Bose–Einstein condensation,[22] and magnon lasers.[23,24]

Among various approaches, PhC nanocavities stand out as a powerful platform for enhancing light–matter interactions by simultaneously achieving high *Q* factors and diffraction-limited mode volumes.[25] However, the fabrication of monocrystalline YIG-based PhC nanocavities remains extremely challenging. Despite numerous and attractive theoretical proposals,[26–28] experimental realizations of monocrystalline YIG-PhCs-based devices have been scarcely reported for nearly three decades, underscoring the difficulty of nanostructure fabrication in this material system.[29]

Here, we challenge the *status quo* and demonstrate photonic crystal defect nanocavities based on monocrystalline YIG. The fabricated nanocavities exhibit strong optical confinement accompanying sharp optical resonances, with a $Q$ factor reaching ~1,800 and a mode volume $V \sim 1.1(\lambda/n)^3$, yielding a remarkably high $Q/V \sim 10^3$. We find that further optimization of the cavity geometry can readily improve the experimental $Q$ factor by an order of magnitude. The YIG-based PhC nanocavities developed in this study will be a novel testbed for exploring light–magnetism interactions under tight spatial confinement.

## 2. Results and Discussion

### 2.1. Fabrication Methodology of YIG PhC Nanocavities

In the course of realizing YIG PhC nanocavities, we have developed two key technologies: monocrystalline YIG thin films on glass and their fine nanopatterning. **Figure 1** conceptually illustrates these technologies and how they enable high *Q* YIG PhC nanocavities. The first technology, monocrystalline YIG thin films on glass, enables tight light confinement in YIG via air cladding by selective undercutting of the glass layer beneath the PhC. This architecture compensates for the YIG's relatively low refractive index of about 2 and maximizes the index contrast for light confinement. The second technology, high-precision nanostructuring of YIG, is crucial for achieving high *Q* factor optical resonances by suppressing light leakage due to fabrication imperfections. The absence of these two technologies has so far precluded the realization of high *Q* YIG PhC nanocavities.

The most common approach to obtaining monocrystalline YIG thin films is epitaxial growth on gadolinium gallium garnet (GGG) substrates,[30] which exhibit a small lattice mismatch with YIG. However, this material combination prevents isolating the YIG layer from the substrate because the two materials share similar chemical properties. Alternative approaches such as sputter deposition,[31] metal organic deposition,[32] and remote epitaxy have also been attempted,[33] but all result in degraded crystallinity.

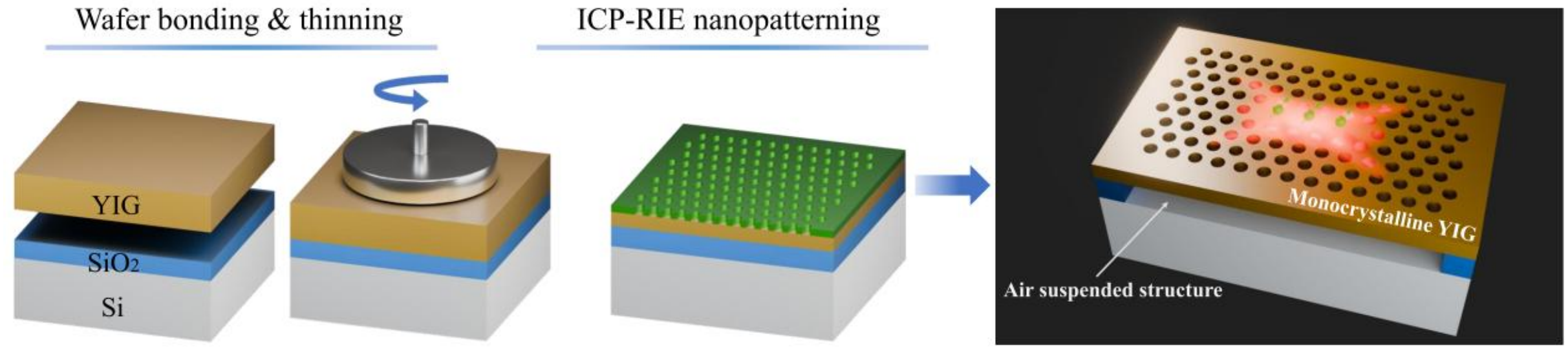

**Figure 1**. Schematic diagram illustrating the concept of this study.

To overcome the existing difficulties, we developed a monocrystalline YIG-on-insulator (YIGOI) structure by a bonding-and-thinning approach[34,35] based on wafer bonding, mechanical polishing, and dry etching.

YIG is chemically and physically resistant to etching and its precise nanopatterning has long been considered extremely difficult. Various approaches have been examined in previous attempts, such as self-assembled templating,[36] lift-off,[37] focused ion beam (FIB) milling and reactive ion etching (RIE) using Ar/$BCl_3$ gas mixtures.[38] However, faithful transfer of the designed patterns with smooth etched sidewalls has remained challenging. In this work, we developed high precision plasma dry etching of YIG using an inductively coupled plasma (ICP) -RIE system with Ar plasma. This approach allows for wafer-scale parallel fabrication while mitigating process damage. In the following, we will discuss the details of the developed fabrication processes.

### 2.2. Fabrication Results

**Figure 2**(a) shows the fabrication process flow of the YIGOI structure developed in this work. A monocrystalline Bi:YIG wafer (thickness: 225 μm; GLB, GRANOPT), grown by liquid-phase epitaxy, was used as the starting material.

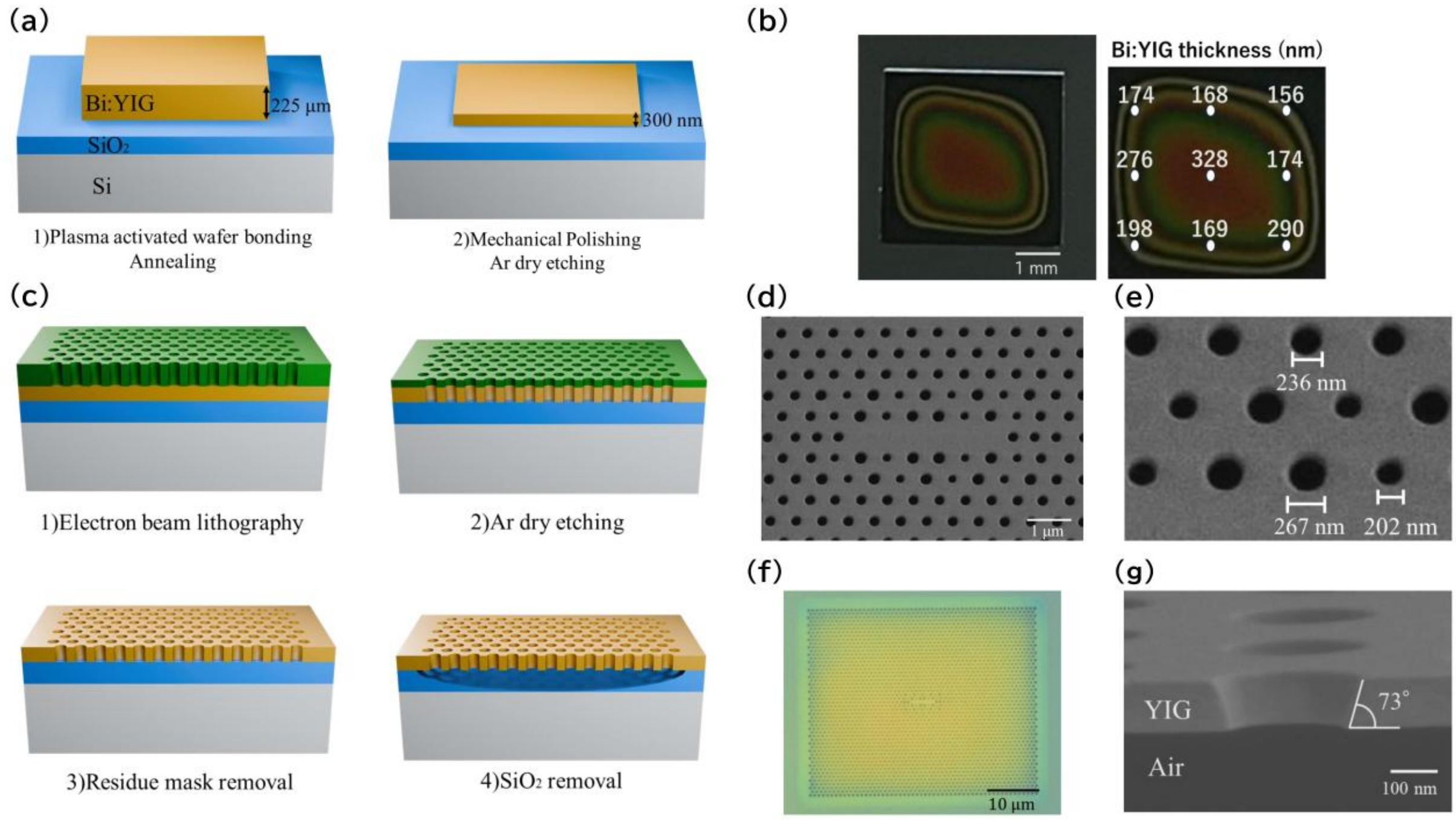

**Figure 2**. Fabrication procedure and images of the fabricated devices. (a) Fabrication flow of the YIGOI substrate. (b) Macroscopic optical image of a YIGOI substrate. The dots and numbers in the enlarged view indicate the thickness of the Bi:YIG layer at the corresponding locations. (c) Fabrication flow of a YIG-based PhC defect nanocavity. (d) Scanning electron microscope (SEM) image of a fabricated PhC nanocavity. (e) Magnified image of (d). (f) Optical microscope image of the fabricated PhC nanocavity. (g) Cross-sectional SEM image of an air-suspended YIG-based PhC.

Bi-doped YIG was chosen because it exhibits a larger MO effect and a higher refractive index ($n = 2.37$) than undoped YIG, making it advantageous for optical applications. The surface of the Bi:YIG wafer prior to bonding showed an average roughness of Ra = 0.10 nm, as characterized by atomic force microscopy (AFM). This wafer and a thermally oxidized Si substrate ($SiO_2$: 2 µm, Si: 625 µm) were cleaned with isopropanol and subsequently bonded using plasma-activated hydrophilic wafer bonding. The bonded wafers were annealed to strengthen adhesion through a dehydration reaction. The YIG layer was then thinned to about 10 µm by mechanical polishing, followed by chemical mechanical polishing (CMP) to smooth the surface. Finally, Ar-based plasma dry etching was applied to further reduce the YIG thickness to approximately 200 - 300 nm, which is suitable for nanophotonics applications in the telecommunication wavelength band. The surface roughness of the fabricated YIGOI substrate was characterized by AFM, yielding Ra = 0.13 nm.

Figure 2(b) presents a photograph of a fabricated YIGOI substrate and the measured thickness distribution of the YIG layer. The local thicknesses were measured using white-light interferometric profilometry. The film thickness is close to the target value of 300 nm around the center of the substrate. Thickness nonuniformity induces a color variation in the film. The thickness variation is relatively gradual around the center of the substrate and measured to be only 0.12 nm/μm, which negligibly impacts on the fabrication of the PhC nanocavities. The thickness nonuniformity predominantly originates from the polishing step, which can be further minimized by optimizing the polishing apparatus. The bonding-and-thinning approach developed here relies purely on physical rather than chemical mechanisms. Therefore, the process is in essence material-independent and possibly applicable to other crystalline materials whose thin films cannot be epitaxially grown on desired substrates.

Next, we discuss the nanopatterning process for PhCs on YIGOI. Figure 2(c) summarizes the developed process flow. We performed electron-beam lithography followed by ICP-RIE with Ar gas and selective etching of the glass layer. In the first step, a thick electron-beam resist layer (ZEP520A, ZEON Corporation) was formed on a YIGOI substrate by two successive spin-coating steps. A conductive polymer (Espacer 300Z, Resonac) was then applied to mitigate charging during exposure, followed by electron-beam lithography to define the PhC nanocavity pattern. Subsequently, the resist pattern was transferred into the YIG layer using Ar-based ICP-RIE. The etching conditions were an ICP power of 15 W, a bias power of 100 W, and a chamber pressure of 0.1 Pa. We found that a higher pressure degrades nanofabrication quality. Therefore, the process was carried out at the lower pressure limit of the etching chamber. After dry etching, the residual etching mask was removed, and finally, the $SiO_2$ sacrificial layer was removed using hydrofluoric acid to obtain air-clad YIG nanocavities.

Figure 2(d) and (e) show top-view SEM images of the fabricated YIG photonic crystal nanocavities, which are L5-type defect nanocavities based on a triangular-lattice air-hole PhC designed with a lattice constant of $a = 572$ nm. The air-hole positions near the defect region were modulated, following the design reported by Kuramochi *et al.*[39] For the ease of optical measurements, a double-periodic modulation of the air-hole radius was also introduced around the defect region.[40] From the pictures, one can confirm that the circular air holes with a diameter of 236 nm are faithfully patterned on the YIG layer. The magnified view in Figure 2(e) further confirms the high-quality fabrication. Indeed, the air-hole radius modulation was precisely reproduced in the YIG membrane. The realized amplitude of the radius modification was measured to be 16 nm, which is close to the designed value of $0.03a = 17.2$ nm.

Figure 2(f) shows an optical microscope image of the fabricated PhC cavity, which confirms that the YIG layer is homogeneously air-suspended without any noticeable damage. Figure 2(g) displays a cross-sectional SEM image of a PhC nanocavity after mechanical cleaving. The sidewall angle of the air holes was measured to be 73°. Both the cavity surface and the air-hole sidewalls appear smooth, indicating that no significant surface roughness was introduced during processing. The tilted sidewalls are in general detrimental to optical confinement but their impact is not critical in the current nanocavity design as we will discuss later. We envision that optimizing etching conditions will improve the verticality of the sidewalls.

### 2.3. Optical Characterization

Now, the fabricated nanocavities were optically characterized by micro-spectroscopic cross-polarized reflection measurements. Light from a white lamp was normally incident on a nanocavity through an objective lens, and reflected light was collected with the same lens and analyzed using a spectrometer. **Figure 3**(a)-(c) show measured reflection spectra for three different samples defined with different lattice constants or YIG slab thicknesses.

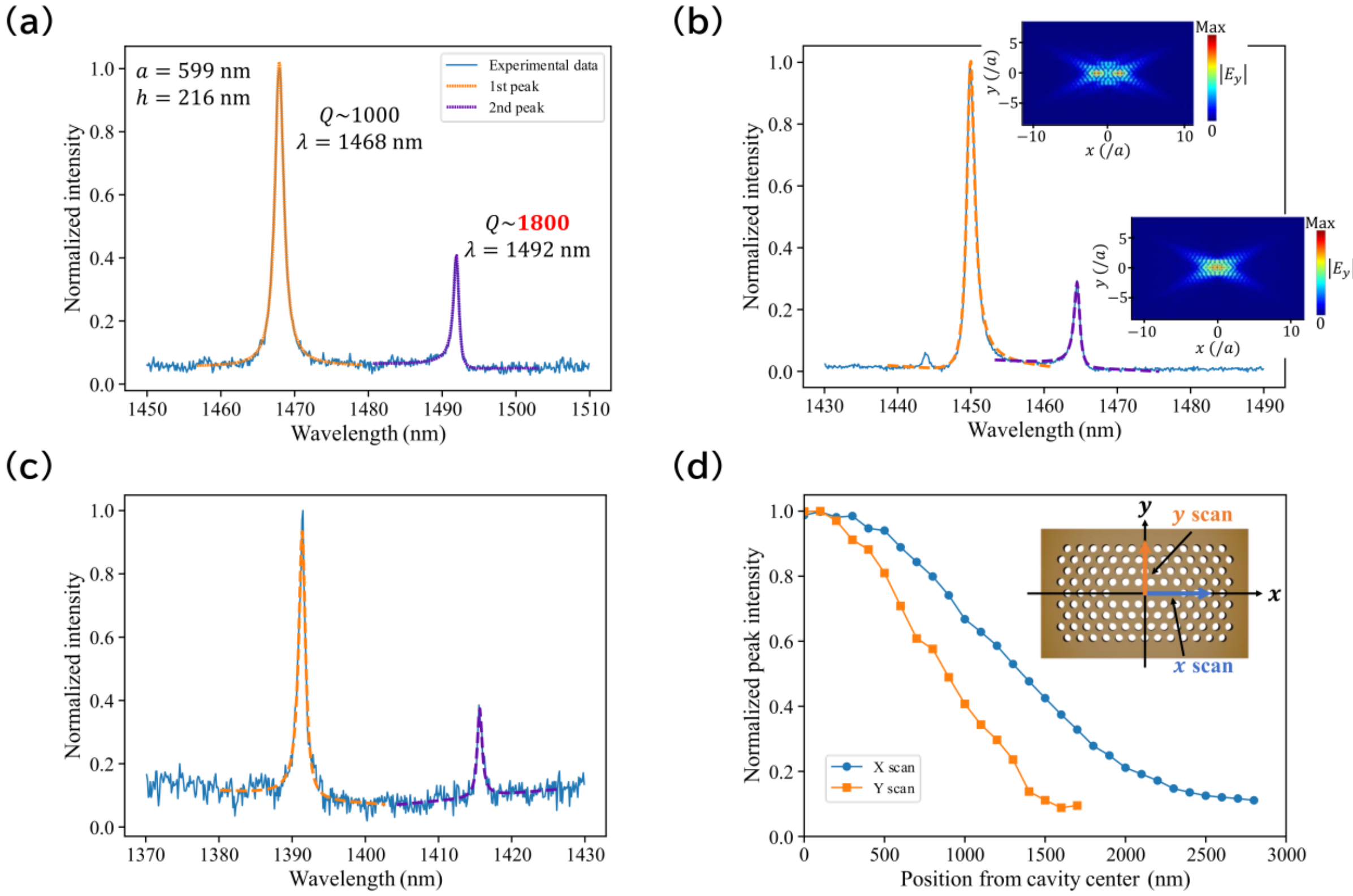

**Figure 3**. Optical characterization of the fabricated nanocavities. Reflection spectra and Fano fitting for cavities with (a) $a = 599$ nm, $h = 216$ nm, (b) $a = 599$ nm, $h = 201$ nm and (c) $a = 585$ nm, $h = 218$ nm. The insets in (b) show the field profiles of the corresponding cavity modes. (d) Spatial dependence of the resonance peak intensity for the first peak in (b); the inset shows scanning directions from the cavity center.

Two distinct peaks were clearly observed in all three samples. The first peak on the shorter-wavelength side exhibits broader resonances, while the second peak at a longer wavelength is narrower. Finite difference time domain (FDTD) simulations reveal that the first (second) resonance corresponds to the first excited (fundamental) TE polarized cavity mode supported in the PhC nanocavity. The corresponding field profiles of these cavity modes are shown in the inset of Figure 3(b). Fano fitting was applied to each spectrum as overlaid in the spectra. For the sample with a lattice constant of $a = 599$ nm and a YIG slab thickness of $h = 216$ nm in Figure 3(a), the resonance peak at $\lambda = 1492$ nm supports a $Q$ of $\sim 1{,}800$. Given that the design $Q$ factor is limited to ~ 2,900 due to the application of the double-periodic modulation, the experimentally measured $Q$ factor comparable to the designed value validates the faithful nanocavity fabrication in the current experiment. From the discrepancy between the design ($Q_{des}$) and experimental ($Q_{exp}$) values, we can deduce the fabrication-limited $Q$ factor to be $Q_{fab} = \left(Q_{exp}^{-1} - Q_{des}^{-1}\right)^{-1} = 9{,}900$. Figure 3(b) and (c) present how the spectra are modified with varying $h$ or $a$. Decreasing $h$ shifts the resonance peaks toward shorter wavelengths and deteriorates the $Q$ factor. These observations can be attributed to a reduction in the effective refractive index of the PhC slab. Meanwhile, the reduction of $a$ only shifts the resonance wavelength without significantly reducing the $Q$ factor. These observations suggest that careful control over $h$ is crucial to attain high $Q$ factor in the current samples.

We further investigated spatial localization of the cavity resonance at 1450 nm observed in Figure 3(b). Figure 3(d) shows the measurement configuration and results. The nanocavity was scanned relatively to the measurement spot under the objective lens while tracking reflection spectrum. The reflection peak intensity is maximized near the cavity center and rapidly decreases as moving away from the center, confirming the spatial localization of the cavity mode within a small region extending only 2 - 3 μm. The intensity drop is more gradual along the x-axis than along the y-axis, reflecting the elongated geometry of the L5 cavity.

### 2.4. Structural Parameter Dependence

The influence of slab thickness $h$ on the optical confinement in the YIG PhC nanocavities is investigated in more detail using the natural thickness variation of the YIG thin film. **Figure 4**(a) and (b) respectively show the measured resonance wavelengths and $Q$ factors as functions of $h$. In this experiment, all samples share the identical structural parameters including the lattice constant ($a = 599$ nm), except for the slab thickness.

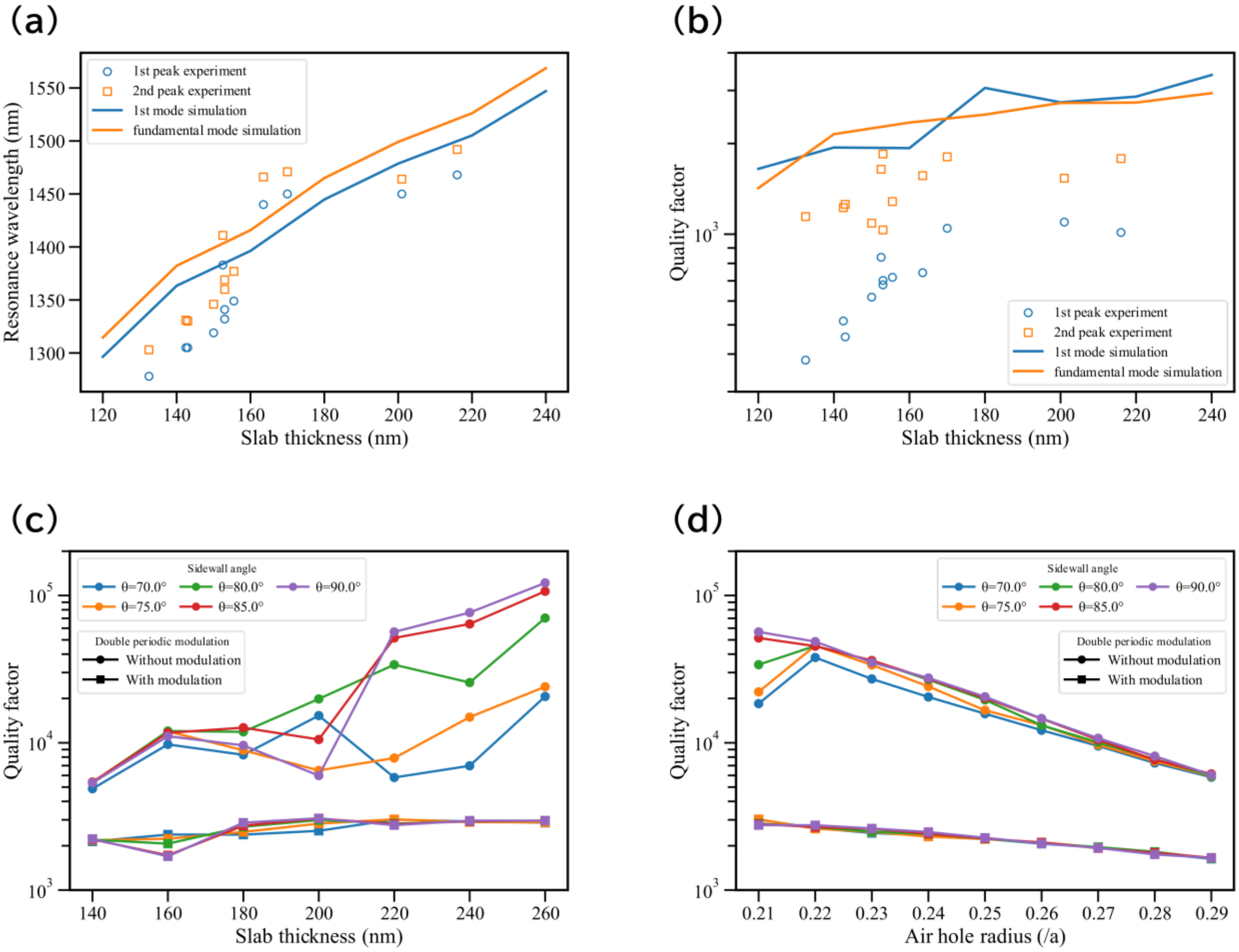

**Figure 4.** (a,b) Experimentally measured and numerically simulated resonance wavelengths and $Q$ factors as a function of slab thickness $h$. The simulation parameters used in these plots are $a = 599$ nm, $r = 0.21a$, $\theta = 73°$ and double-period modulation amplitude $0.03a$. (c) Sidewall angle dependences of computed $Q$ factors for the fundamental mode when varying $h$, while fixing $r = 0.21a$. (d) Same with (c) but when varying air-hole radius $r$ while fixing $h = 220$ nm. The two series of data in each panel are of the structures excluding and including the double-periodic air-hole modulation with an amplitude $0.03a$. All results are obtained for the PhCs designed with a lattice constant of $a = 599$ nm.

As shown in the plots, both two resonance peaks found in a single nanocavity shift toward shorter wavelengths as $h$ decreased. Concomitantly, the $Q$ factors decrease with decreasing $h$. By thinning $h$ from 216 nm to 133 nm, the resonance wavelength of the first peak changes from 1468 nm to 1278 nm, and that of the second peak from 1492 nm to 1303 nm, while the separation between the two peaks remains nearly constant at approximately 25 nm. More importantly, decreasing $h$ significantly impacts on the $Q$ factors. Varying $h$ from 216 nm to 133 nm, the $Q$ factor of the first resonance peak decreases from 1,000 to 400, and that of the second peak changes from 1,800 to 1,100. The larger impact on the first resonance's $Q$ factor may arise from its weaker optical confinement, which is more influenced by the reduction of the effective refractive index of the PhC slab induced by thinning $h$. In the plots, a noticeable sample-to-sample variation is observed, which can be attributed to the random fluctuation of air-hole radii and positions introduced during the fabrication.

Electromagnetic field simulations were performed to analyze how device geometry influences optical confinement. We performed FDTD simulations based on the structural parameters retrieved from the SEM images: $a = 599$ nm, air-hole radii $r = 0.21a$, sidewall angle $\theta = 73°$, double-period modulation amplitude $0.03a$. The simulation reproduced the fundamental and the first-order TE-like modes in the wavelength range of interest. The simulated resonance wavelengths and $Q$ factors of these modes when varying $h$ are plotted as solid lines in Figure 4(a) and (b). Both modes showed a monotonic blue shift with decreasing $h$. The calculated $Q$ factors also decreased slightly as $h$ reduced, showing a similar trend to the experimental results. However, the absolute $Q$ factors were lower in the experiments, likely due to random variations in air-hole position and radius that are not considered in the simulations.

We continued numerical simulations to understand the impact of the tilted sidewall angle $\theta$ on $Q$ factor, together with the impact of slab thickness $h$ and the air-hole radius $r$. Figure 4(c) and (d) show the simulated $Q$ factors of the fundamental cavity mode for different $\theta$ as a function of $h$ and $r$, respectively. In the plots, both the results for the cavities with and without the double-periodic modulation are displayed. In Figure 4(c), $r = 0.21a$ was fixed, while in Figure 4(d), $h = 220$ nm was fixed. As a general tendency, reducing $h$ obscures the influence of the tapered air holes, as confirmed by the converging behavior of $Q$ factors computed for different $\theta$ at $h$ = 140 nm, at which $Q$ becomes ~ 5,000 for all the designs. Thinner $h$ will also help the fabrication process as only requiring shallow etching of the YIG layer. On the other hand, a thicker slab is preferable for increasing the effective refractive index of the PhC and thus for strengthening the optical confinement via the index contrast. Indeed, $Q$ factor exceeds 10,000 even for $\theta = 70°$ when $h = 200$ nm. In short, careful optimization of the slab thickness for both the design and fabrication perspectives is indispensable for achieving high $Q$ factors. A similar optimization must be considered for $r$. Figure 4(d) indicates that smaller values of $r$ lead to higher simulated $Q$ factors. This tendency can be attributed to the higher effective refractive index in designs with smaller $r$. However, the smaller air holes are more difficult to perforate in the experiment. Interestingly, the influence of tilted sidewalls is less pronounced when increasing $r$, which may be associated with the reduced effective refractive index. To experimentally achieve high $Q$ factors, one needs to carefully optimize $r$ as well.

In the cavity designs used for the experiments, the double-periodic modulation predominantly determines the theoretical $Q$ factors to be around 2,000 ~ 3,000. While this modulation significantly enhances the signal-to-noise ratio (S/N) in the measurements via strengthening the vertical radiation, it clamps the experimental $Q$ factors below 1,800 and obscures the achievable $Q$ factor with the current fabrication quality. If we eliminate the modulation, the present cavity geometry ($r = 0.21a, h = 220$ nm, $\theta = 73°$) supports a much higher design $Q$ of $> 10^4$. Therefore, a substantial improvement in $Q$ is expected simply by utilizing a measurement configuration that does not rely on the vertical cavity leakage, such as those through waveguide coupling.

## 3. Conclusion

In this work, we demonstrated PhC defect nanocavities based on monocrystalline Bi:YIG and observed cavity resonances with high $Q$ factors up to 1,800 in the telecommunication wavelengths. The measured $Q$ factors were predominantly governed by design and can be improved by nearly an order-of-magnitude through geometrical optimization. These results were enabled by the development of a monocrystalline YIG-on-insulator platform and its precise nanopatterning. The YIGOI substrate established here is expected to advance the study of magnonics based on YIG.[41] The new architecture can isolate monocrystalline YIG thin films from GGG substrates, which are known to deteriorate spin coherence in YIG at cryogenic temperatures.[42] Beyond YIG-based magneto-optical nanophotonics, the present platform provides a basis for engineering both optical and magnonic modes in monocrystalline YIG. Such nanostructures could improve photon–magnon mode overlap in cavity optomagnonics, an essential step toward integrated microwave-to-optical quantum transducers. The ability to form periodic nanostructures in low-damping YIG may also enable optomagnonic and magnonic crystals with engineered spin-wave band structures, offering opportunities for reconfigurable magnonic devices,[43] topological magnonic transport,[44] and quantum magnonic interfaces. Finally, the fabrication techniques discussed here are highly versatile as they mostly rely on physical processes and thus can be applicable to many other materials.[45,46] Our results not only advance YIG-based MO nanophotonics but also pave the way for photonic applications of other materials that are challenging to microfabricate.

Acknowledgements

We would like to thank H. Otsuki, H. Matsukiyo, M. Nishioka, S. Ishida, R. Hisatomi, K. Usami and A. I. Musorin for their technical support and fruitful discussions. This work was supported by JST FOREST (JPMJFR213F), JST CREST (JPMJCR19T1) and KAKENHI (25K01697, 24K17582), Iketani Foundation, Nippon Sheet Glass Foundation, and JST SPRING Program (JPMJSP2123).

**References**

(1) Bi, L. Materials for Nonreciprocal Photonics. *MRS Bull.* **2018**, *43* (6), 408–412. https://doi.org/10.1557/mrs.2018.120.

(2) Schmidt, G.; Hauser, C.; Trempler, P.; Paleschke, M.; Papaioannou, E. T. Ultra Thin Films of Yttrium Iron Garnet with Very Low Damping: A Review. *Phys. Status Solidi Basic Res.* **2020**, *257* (7). https://doi.org/10.1002/pssb.201900644.

(3) Srinivasan, K.; Stadler, B. J. H. Review of Integrated Magneto-Optical Isolators with Rare-Earth Iron Garnets for Polarization Diverse and Magnet-Free Isolation in Silicon Photonics [Invited]. *Opt. Mater. Express* **2022**, *12* (2), 697. https://doi.org/10.1364/ome.447398.

(4) Schloemann, E. F. Circulators for Microwave and Millimeter-Wave Integrated Circuits. *Proc. IEEE* **1988**, *76* (2), 188–200. https://doi.org/10.1109/5.4394.

(5) Serga, A. A.; Chumak, A. V.; Hillebrands, B. YIG Magnonics. *J. Phys. D. Appl. Phys.* **2010**, *43* (26). https://doi.org/10.1088/0022-3727/43/26/264002.

(6) Yang, W.; Qin, J.; Long, J.; Yan, W.; Yang, Y.; Li, C.; Li, E.; Hu, J.; Deng, L.; Du, Q.; Bi, L. A Self-Biased Non-Reciprocal Magnetic Metasurface for Bidirectional Phase Modulation. *Nat. Electron.* **2023**, *6* (3), 225–234. https://doi.org/10.1038/s41928-023-00936-w.

(7) Wang, Z.; Chong, Y.; Joannopoulos, J. D.; Soljačić, M. Observation of Unidirectional Backscattering-Immune Topological Electromagnetic States. *Nature* **2009**, *461* (7265), 772–775. https://doi.org/10.1038/nature08293.

(8) Tabuchi, Y.; Ishino, S.; Ishikawa, T.; Yamazaki, R.; Usami, K.; Nakamura, Y. Hybridizing Ferromagnetic Magnons and Microwave Photons in the Quantum Limit. *Phys. Rev. Lett.* **2014**, *113* (8), 1–5. https://doi.org/10.1103/PhysRevLett.113.083603.

(9) Inoue, M.; Arai, K.; Fujii, T.; Abe, M. One-Dimensional Magnetophotonic Crystals. *J. Appl. Phys.* **1999**, *85* (8), 5768–5770. https://doi.org/10.1063/1.370120.

(10) Chin, J. Y.; Steinle, T.; Wehlus, T.; Dregely, D.; Weiss, T.; Belotelov, V. I.; Stritzker, B.; Giessen, H. Nonreciprocal Plasmonics Enables Giant Enhancement of Thin-Film Faraday Rotation. *Nat. Commun.* **2013**, *4*. https://doi.org/10.1038/ncomms2609.

(11) Borovkova, O. V.; Hashim, H.; Ignatyeva, D. O.; Kozhaev, M. A.; Kalish, A. N.; Dagesyan, S. A.; Shaposhnikov, A. N.; Berzhansky, V. N.; Achanta, V. G.; Panina, L. V.; Zvezdin, A. K.; Belotelov, V. I. Magnetoplasmonic Structures with Broken Spatial Symmetry for Light Control at Normal Incidence. *Phys. Rev. B* **2020**, *102* (8), 1–7. https://doi.org/10.1103/PhysRevB.102.081405.

(12) Yang, W.; Liu, Q.; Wang, H.; Chen, Y.; Yang, R.; Xia, S.; Luo, Y.; Deng, L.; Qin, J.; Duan, H.; Bi, L. Observation of Optical Gyromagnetic Properties in a Magneto-Plasmonic Metamaterial. *Nat. Commun.* **2022**, *13* (1), 1–8. https://doi.org/10.1038/s41467-022-29452-9.

(13) Voronov, A. A.; Karki, D.; Ignatyeva, D. O.; Kozhaev, M. A.; Levy, M.; Belotelov, V. I. Magneto-Optics of Subwavelength All-Dielectric Gratings. *Opt. Express* **2020**, *28* (12), 17988. https://doi.org/10.1364/OE.394722.

(14) Ignatyeva, D. O.; Karki, D.; Voronov, A. A.; Kozhaev, M. A.; Krichevsky, D. M.; Chernov, A. I.; Levy, M.; Belotelov, V. I. All-Dielectric Magnetic Metasurface for Advanced Light Control in Dual Polarizations Combined with High-Q Resonances. *Nat. Commun.* **2020**, *11* (1), 5487. https://doi.org/10.1038/s41467-020-19310-x.

(15) Qin, J.; Xia, S.; Yang, W.; Wang, H.; Yan, W.; Yang, Y.; Wei, Z.; Liu, W.; Luo, Y.; Deng, L.; Bi, L. Nanophotonic Devices Based on Magneto-Optical Materials: Recent Developments and Applications. *Nanophotonics* **2022**, *11* (11), 2639–2659. https://doi.org/10.1515/nanoph-2021-0719.

(16) Zhang, X.; Zhu, N.; Zou, C. L.; Tang, H. X. Optomagnonic Whispering Gallery Microresonators. *Phys. Rev. Lett.* **2016**, *117* (12), 1–5. https://doi.org/10.1103/PhysRevLett.117.123605.

(17) Osada, A.; Hisatomi, R.; Noguchi, A.; Tabuchi, Y.; Yamazaki, R.; Usami, K.; Sadgrove, M.; Yalla, R.; Nomura, M.; Nakamura, Y. Cavity Optomagnonics with Spin-Orbit Coupled Photons. *Phys. Rev. Lett.* **2016**, *116* (22), 1–5. https://doi.org/10.1103/PhysRevLett.116.223601.

(18) Haigh, J. A.; Langenfeld, S.; Lambert, N. J.; Baumberg, J. J.; Ramsay, A. J.; Nunnenkamp, A.; Ferguson, A. J. Magneto-Optical Coupling in Whispering-Gallery-Mode Resonators. *Phys. Rev. A - At. Mol. Opt. Phys.* **2015**, *92* (6), 1–7. https://doi.org/10.1103/PhysRevA.92.063845.

(19) Zhu, N.; Zhang, X.; Han, X.; Zou, C. L.; Tang, H. X. Inverse Faraday Effect in an Optomagnonic Waveguide. *Phys. Rev. Appl.* **2022**, *18* (2), 1. https://doi.org/10.1103/PhysRevApplied.18.024046.

(20) Zare Rameshti, B.; Viola Kusminskiy, S.; Haigh, J. A.; Usami, K.; Lachance-Quirion, D.; Nakamura, Y.; Hu, C.-M.; Tang, H. X.; Bauer, G. E. W.; Blanter, Y. M. Cavity Magnonics. *Phys. Rep.* **2022**, *979*, 1–61. https://doi.org/10.1016/j.physrep.2022.06.001.

(21) Hisatomi, R.; Osada, A.; Tabuchi, Y.; Ishikawa, T.; Noguchi, A.; Yamazaki, R.; Usami, K.; Nakamura, Y. Bidirectional Conversion between Microwave and Light via Ferromagnetic Magnons. *Phys. Rev. B* **2016**, *93* (17), 1–13. https://doi.org/10.1103/PhysRevB.93.174427.

(22) Demokritov, S. O.; Demidov, V. E.; Dzyapko, O.; Melkov, G. A.; Serga, A. A.; Hillebrands, B.; Slavin, A. N. Bose–Einstein Condensation of Quasi-Equilibrium Magnons at Room Temperature under Pumping. *Nature* **2006**, *443* (7110), 430–433. https://doi.org/10.1038/nature05117.

(23) He, X.-W.; Wang, Z.-Y.; Han, X.; Zhang, S.; Wang, H.-F. Parametrically Amplified Nonreciprocal Magnon Laser in a Hybrid Cavity Optomagnonical System. *Opt. Express* **2023**, *31* (26), 43506. https://doi.org/10.1364/oe.509918.

(24) Wang, B.; Jia, X.; Lu, X. H.; Xiong, H. PT -Symmetric Magnon Laser in Cavity Optomagnonics. *Phys. Rev. A* **2022**, *105* (5), 1–9. https://doi.org/10.1103/PhysRevA.105.053705.

(25) Akahane, Y.; Asano, T.; Song, B. S.; Noda, S. High-Q Photonic Nanocavity in a Two-Dimensional Photonic Crystal. *Nature* **2003**, *425* (6961), 944–947. https://doi.org/10.1038/nature02063.

(26) Wang, Z.; Fan, S. Magneto-Optical Defects in Two-Dimensional Photonic Crystals. *Appl. Phys. B Lasers Opt.* **2005**, *81* (2–3), 369–375. https://doi.org/10.1007/s00340-005-1846-x.

(27) Christofi, A.; Kawaguchi, Y.; Alù, A.; Khanikaev, A. B. Giant Enhancement of Faraday Rotation Due to Electromagnetically Induced Transparency in All-Dielectric Magneto-Optical Metasurfaces. *Opt. Lett.* **2018**, *43* (8), 1838. https://doi.org/10.1364/ol.43.001838.

(28) Gao, S.; Liu, T.; Iwamoto, S.; Ota, Y. Design of Ultrathin Faraday Rotators Based on All-Dielectric Magneto-Optical Metasurfaces at the Telecommunication Band. *ACS Photonics* **2025**, *12* (12), 6745–6751. https://doi.org/10.1021/acsphotonics.5c01826.

(29) Magdenko, L.; Popova, E.; Vanwolleghem, M.; Pang, C.; Fortuna, F.; Maroutian, T.; Beauvillain, P.; Keller, N.; Dagens, B. Wafer-Scale Fabrication of Magneto-Photonic Structures in Bismuth Iron Garnet Thin Film. *Microelectron. Eng.* **2010**, *87* (11), 2437–2442. https://doi.org/10.1016/j.mee.2010.04.021.

(30) Ignatyeva, D. O.; Krichevsky, D. M.; Belotelov, V. I.; Royer, F.; Dash, S.; Levy, M. All-Dielectric Magneto-Photonic Metasurfaces. *J. Appl. Phys.* **2022**, *132* (10). https://doi.org/10.1063/5.0097607.

(31) Onbasli, M. C.; Goto, T.; Sun, X.; Huynh, N.; Ross, C. A. Integration of Bulk-Quality Thin Film Magneto-Optical Cerium-Doped Yttrium Iron Garnet on Silicon Nitride Photonic Substrates. *Opt. Express* **2014**, *22* (21), 25183. https://doi.org/10.1364/oe.22.025183.

(32) Ishibashi, T.; Kawata, T.; Johansen, T. H.; He, J.; Harada, N.; Sato, K. Magneto-Optical Indicator Garnet Films Grown by Metal-Organic Decomposition Method. *J. Magn. Soc. Japan* **2008**, *32* (2_2), 150–153. https://doi.org/10.3379/msjmag.32.150.

(33) Kum, H. S.; Lee, H.; Kim, S.; Lindemann, S.; Kong, W.; Qiao, K.; Chen, P.; Irwin, J.; Lee, J. H.; Xie, S.; Subramanian, S.; Shim, J.; Bae, S. H.; Choi, C.; Ranno, L.; Seo, S.; Lee, S.; Bauer, J.; Li, H.; Lee, K.; Robinson, J. A.; Ross, C. A.; Schlom, D. G.; Rzchowski, M. S.; Eom, C. B.; Kim, J. Heterogeneous Integration of Single-Crystalline Complex-Oxide Membranes. *Nature* **2020**, *578* (7793), 75–81. https://doi.org/10.1038/s41586-020-1939-z.

(34) Lukin, D. M.; Dory, C.; Guidry, M. A.; Yang, K. Y.; Mishra, S. D.; Trivedi, R.; Radulaski, M.; Sun, S.; Vercruysse, D.; Ahn, G. H.; Vučković, J. 4H-Silicon-Carbide-on-Insulator for Integrated Quantum and Nonlinear Photonics. *Nat. Photonics* **2020**, *14* (5), 330–334. https://doi.org/10.1038/s41566-019-0556-6.

(35) Yang, J.; Van Gasse, K.; Lukin, D. M.; Guidry, M. A.; Ahn, G. H.; White, A. D.; Vučković, J. Titanium:Sapphire-on-Insulator Integrated Lasers and Amplifiers. *Nature* **2024**, *630* (8018), 853–859. https://doi.org/10.1038/s41586-024-07457-2.

(36) Ikezawa, Y.; Nishimura, K.; Uchida, H.; Inoue, M. Preparation of Two-Dimensional Magneto-Photonic Crystals of Bismus Substitute Yttrium Iron Garnet Materials. *J. Magn. Magn. Mater.* **2004**, *272–276* (III), 1690–1691. https://doi.org/10.1016/j.jmmm.2003.12.256.

(37) Li, S.; Zhang, W.; Ding, J.; Pearson, J. E.; Novosad, V.; Hoffmann, A. Epitaxial Patterning of Nanometer-Thick Y3Fe5O12 Films with Low Magnetic Damping. *Nanoscale* **2016**, *8* (1), 388–394. https://doi.org/10.1039/c5nr06808h.

(38) Rashedi, A.; Ebrahimi, M.; Huang, Y.; Rudd, M. J.; Davis, J. P.; Bittencourt, V. A. S. V. Photonic Crystal Cavities Based on Suspended Yttrium Iron Garnet Nanobeams. *Phys. Rev. Appl.* **2025**, *24* (5), 054017. https://doi.org/10.1103/xptl-hx1j.

(39) Kuramochi, E.; Grossman, E.; Nozaki, K.; Takeda, K.; Shinya, A.; Taniyama, H.; Notomi, M. Systematic Hole-Shifting of L-Type Nanocavity with an Ultrahigh Q Factor. *Opt. Lett.* **2014**, *39* (19), 5780. https://doi.org/10.1364/ol.39.005780.

(40) Tran, N. V. Q.; Combrié, S.; De Rossi, A. Directive Emission from High-Q Photonic Crystal Cavities through Band Folding. *Phys. Rev. B - Condens. Matter Mater. Phys.* **2009**, *79* (4), 1–4. https://doi.org/10.1103/PhysRevB.79.041101.

(41) Pintus, P.; Ranzani, L.; Pinna, S.; Huang, D.; Gustafsson, M. V.; Karinou, F.; Casula, G. A.; Shoji, Y.; Takamura, Y.; Mizumoto, T.; Soltani, M.; Bowers, J. E. An Integrated Magneto-Optic Modulator for Cryogenic Applications. *Nat. Electron.* **2022**, *5* (September). https://doi.org/10.1038/s41928-022-00823-w.

(42) Serha, R. O.; Voronov, A. A.; Schmoll, D.; Klingbeil, R.; Knauer, S.; Koraltan, S.; Pribytova, E.; Lindner, M.; Reimann, T.; Dubs, C.; Abert, C.; Verba, R.; Urbánek, M.; Suess, D.; Chumak, A. V. Damping Enhancement in YIG at Millikelvin Temperatures Due to GGG Substrate. *Mater. Today Quantum* **2025**, *5* (December 2024), 100025. https://doi.org/10.1016/j.mtquan.2025.100025.

(43) Krawczyk, M.; Grundler, D. Review and Prospects of Magnonic Crystals and Devices with Reprogrammable Band Structure. *J. Phys. Condens. Matter* **2014**, *26* (12). https://doi.org/10.1088/0953-8984/26/12/123202.

(44) Shindou, R.; Matsumoto, R.; Murakami, S.; Ohe, J. I. Topological Chiral Magnonic Edge Mode in a Magnonic Crystal. *Phys. Rev. B - Condens. Matter Mater. Phys.* **2013**, *87* (17), 2–12. https://doi.org/10.1103/PhysRevB.87.174427.

(45) Liang, H.; Luo, R.; He, Y.; Jiang, H.; Lin, Q. High-Quality Lithium Niobate Photonic Crystal Nanocavities. *Optica* **2017**, *4* (10), 1251. https://doi.org/10.1364/OPTICA.4.001251.

(46) Li, M.; Liang, H.; Luo, R.; He, Y.; Lin, Q. High-Q 2D Lithium Niobate Photonic Crystal Slab Nanoresonators. *Laser Photonics Rev.* **2019**, *13* (5). https://doi.org/10.1002/lpor.201800228.